\catcode`\@=11

\@addtoreset{equation}{section}

\def\@normalsize{\@setsize\normalsize{15pt}\xiipt\@xiipt
\abovedisplayskip 14pt plus3pt minus3pt%
\belowdisplayskip \abovedisplayskip
\abovedisplayshortskip  \z@ plus3pt%
\belowdisplayshortskip  7pt plus3.5pt minus0pt}

\def\small{\@setsize\small{13.6pt}\xipt\@xipt
\abovedisplayskip 13pt plus3pt minus3pt%
\belowdisplayskip \abovedisplayskip
\abovedisplayshortskip  \z@ plus3pt%
\belowdisplayshortskip  7pt plus3.5pt minus0pt

\def\@listi{\parsep 4.5pt plus 2pt minus 1pt
            \itemsep \parsep
            \topsep 9pt plus 3pt minus 3pt}}

\def\underline#1{\relax\ifmmode\@@underline#1\else
        $\@@underline{\hbox{#1}}$\relax\fi}
\@twosidetrue

\catcode`@=12

\catcode`\@=11

\catcode`\@=12

\relax

\def\figcap{\section*{Figure Captions\markboth
        {FIGURECAPTIONS}{FIGURECAPTIONS}}\list
        {Fig. \arabic{enumi}:\hfill}{\settowidth\labelwidth{Fig. 999:}
        \leftmargin\labelwidth
        \advance\leftmargin\labelsep\usecounter{enumi}}}
 \relax
\def\tablecap{\section*{Table Captions\markboth
        {TABLECAPTIONS}{TABLECAPTIONS}}\list
        {Table \arabic{enumi}:\hfill}{\settowidth\labelwidth{Table 999:}
        \leftmargin\labelwidth
        \advance\leftmargin\labelsep\usecounter{enumi}}}
 \relax
\def\reflist{\subsubsection*{References\markboth
        {REFLIST}{REFLIST}}\list
        {[\arabic{enumi}]\hfill}{\settowidth\labelwidth{[999]}
        \leftmargin\labelwidth
        \advance\leftmargin\labelsep\usecounter{enumi}}}
 \relax

\catcode`\@=11

\def\FERMIPUB{}

\def\ps@headings{\def\@oddfoot{}\def\@evenfoot{}
\def\@oddhead{\hbox{}\hfill
        \makebox[.5\textwidth]{\raggedright\ignorespaces --\thepage{}--
        \hfill {\rm FERMILAB--Pub--\FERMIPUB}}}
\def\@evenhead{\@oddhead}
\def\subsectionmark##1{\markboth{##1}{}}
}

\ps@headings

\relax

\newskip\humongous \humongous=0pt plus 1000pt minus 1000pt

\newif\ifdtup

\def\beq{\begin{equation}}
\def\eeq{\end{equation}}

\def\beqn{\begin{eqnarray}}
\def\eeqn{\end{eqnarray}}
\relax

\def\dotx{\dotx{\dot\overline{x}}}

\documentstyle[art12]{article}
\def\today{\number\day\space
     \ifcase\month\or
       January\or February\or March\or April\or May\or June\or
       July\or August\or September\or October\or November\or December\fi
     \space\number\year}
\newcommand{\degr}{^0}
\begin{document}
\begin{titlepage}
\begin{flushright}
Z\"urich University ZU-TH 21/96\\
\end{flushright}
\vfill
\vskip 1cm
\begin{center}
{\large\bf NEUTRINOS AND DARK MATTER IN GALACTIC HALOS
\footnote{
Invited lecture to appear in the proceedings of the Zuoz Summer School
on Physics with neutrinos
(Zuoz, 4-10 August 1996)}}\\
\vskip 0.5cm
{\bf Philippe~Jetzer}\\
\vskip 0.5cm
Paul Scherrer Institute, Laboratory for Astrophysics,
CH-5232 Villigen PSI, and\\
Institute of Theoretical Physics, University of Z\"urich,
Winterthurerstrasse 190,\\
CH-8057 Z\"urich, Switzerland\\
\end{center}
\vskip 0.5 cm
\begin{center}
Abstract
\end{center}
\begin{quote}
One of the most important problems in astrophysics concerns the nature
of the dark matter in galactic halos, whose presence is 
implied mainly by the 
observed flat rotation curves in spiral galaxies.
Due to the Pauli exclusion principle it can be shown that
neutrinos cannot be 
a major constituent of the halo dark matter.
As far as cold dark matter is concerned there might be a discrepancy
between the results of the N-body simulations
and the measured rotation curves for dwarf 
galaxies. A fact this, if confirmed, which would exclude 
cold dark matter  as a viable candidate for the halo dark matter.

In the framework of a baryonic scenario the most plausible 
candidates are brown or white dwarfs and cold molecular clouds
(mainly of $H_2$).
The former can be detected with the ongoing microlensing experiments.
In fact,
the French collaboration EROS and the American-Australian
collaboration MACHO have reported the observation of 
altogether $\sim$ 10 microlensing
events by monitoring during several years the brightness of millions
of stars in the Large Magellanic Cloud. 
In particular, the MACHO team announced the discovery of 
8 microlensing candidates by analysing their first 2 years
of observations. This would imply that the halo dark matter
fraction in form of MACHOs 
(Massive Astrophysical Compact Halo Objects)
is of the order of 45-50\% assuming a standard spherical halo model.
The most accurate way
to get information on the mass of the MACHOs
is to use the method of mass moments, which leads
to an average mass of 0.27$M_{\odot}$.
\end{quote}
\vfill
\end{titlepage}
\newpage
\section{Introduction}

One of the most important problems in astrophysics concerns
the nature of the dark matter in galactic halos, whose presence is 
implied by the observed flat rotation curves in spiral galaxies
\cite{kn:Faber,kn:Trimble}, the X-ray diffuse emission
in elliptical galaxies as well as by the dynamics of galaxy
clusters. Primordial nucleosynthesis entails that most of the baryonic 
matter in the Universe is nonluminous, and such an amount of dark matter falls
suspiciously close to that required by the rotation curves.
Surely, the standard model of elementary particle forces can hardly
be viewed as the ultimate theory and all the attempts in that
direction invariably call for new particles. Hence, the idea of
nonbaryonic dark matter naturally enters the realm of cosmolgy
and may help in the understanding of the process of galaxy
formation and clustering of galaxies.  

The problem of dark matter started already with the pioneering work
of Oort \cite{kn:Oort} in 1932 and Zwicky \cite{kn:Zwicky} in 1933
and its mistery is still not solved.
Actually, there are several dark matter problems on different scales
ranging from the solar neighbourhood, galactic halos, cluster of galaxies
to cosmological scales. Dark matter is also needed to understand
the formation of large scale structures in the universe.

Many candidates have been proposed, either baryonic or not,
to explain dark matter. It is beyond the scope of this lecture to go
through all of these candidates.
Here, we restrict ourself to the dark matter
in galactic halos, and in particular in the halo of our own
Galaxy.
First, we review the evidence for dark matter near the Sun and
in the halo of spiral galaxies. In Section 3 we discuss
the constraint due to the Pauli exclusion principle on neutrinos
as a candidate for halo dark matter. Based on that 
argument neutrinos can practically be excluded as a major
constituent of the dark galactic halos.
In Section 4 we present the baryonic candidates
and in Section 5 we elaborate in some detail on the detection
of MACHOs (Massive Astrophysical Compact Halo Objects) through
microlensing as well as on the most recent observations.
Section 6 is devoted to a scenario in which part of the dark matter
is in the form of cold molecular clouds (mainly of $H_2$).

\section{Evidence for dark matter}

In this Section we briefly outline the evidence
for dark matter in the solar neighbourhood and in the halos of spiral
galaxies. Moreover, we discuss also the total mass of our Galaxy, which can be
inferred from studies of the proper motion of the satellites of
the Milky Way.

\subsection{Dark matter near the Sun}
The local mass density \cite{kn:Binney}
in main sequence and giant stars, stellar
remnants (directly observed or inferred from models of galactic
and stellar evolution), gas and dust yields a lower limit to the total
density of $\rho \simeq 0.114~M_{\odot}$ pc$^{-3}$.
Correspondingly, one finds a mass-to-light ratio of
\begin{equation}
\Upsilon \simeq 1.7 \Upsilon_{\odot}
\end{equation}
($\Upsilon_{\odot}=M_{\odot}/L_{\odot}$, where $M_{\odot}$ is the
mass and $L_{\odot}$ the luminosity of the Sun, respectively).
The local mass density is determined from carefully selected star
samples by analyzing the velocity dispersion and density profile
in the direction normal to the galactic plane.
This yields a total density $\rho = 0.18 \pm 0.03 ~M_{\odot}$ pc$^{-3}$
for the local matter, or equivalently
\begin{equation}
\Upsilon \simeq 2.7 \Upsilon_{\odot}~.
\end{equation}
Therefore, at least 0.03 $M_{\odot}$ pc$^{-3}$ is the contribution from
dark matter. The presence of disk dark matter has long been suspected
\cite{kn:Oort} and it is most likely baryonic.

Recently, at least 8 brown dwarfs have been detected within
a distance of about 100 light years from the Sun. One of these
brown dwarfs is about 70 million years old and has an estimated mass 
of $0.065 ~M_{\odot}$. Moreover, some brown dwarfs have been found
orbiting brighter compagnons, and other as free flying in the Pleiades
cluster.
It is still premature to infer on their contribution to the local
dark matter, although it is plausible that they may make up
an important fraction, if not all.
 
\subsection{Rotation curves of spiral galaxies}

The best evidence for dark matter in galaxies comes from the rotation
curves of spirals.
Measurements of the rotation velocity $v_{rot}$ of stars up to the 
visible edge 
of the spiral galaxies and of $HI$ gas in the disk beyond the
optical radius (by measuring the 
Doppler shift in the 21-cm line) imply that $v_{rot} \approx$ constant
out to very large distances, rather than to show a Keplerian falloff.
These observations started around 1970 \cite{kn:Rubin}, thanks
to the improved sensitivity in both optical and 21-cm bands.
By now there are observations for over thousand 
spiral galaxies with reliable
rotation curves out to large radii \cite{kn:Persic}. In almost all of them
the rotation curve is flat or slowly rising out to the last measured
point. Very few galaxies show falling rotation curves and those
that do either fall less rapidly than Keplerian have nearby companions
that may perturb the velocity field or have large spheroids
that may increase the rotation velocity near the centre.

One of the best exemple for the measurement of the rotation velocity
is the spiral galaxy NGC 3198 \cite{kn:vanAlbada} (see Fig. 1).
Its halo density can be fitted by the following
formula
\begin{equation}
\rho(r) = \frac{\rho_0}{1+(r/a)^{\gamma}}~,
\end{equation}
where $\rho_0 = 0.013 h_0^2~M_{\odot}$ pc$^{-3}$ ($h_0$ being the
Hubble constant in units of $H_0 =100 h_0$ km s$^{-1}$ kpc$^{-1}$, and
$0.4 \leq h_0 \leq 1$), $a = 6.4 h_0^{-1}$ kpc, and $\gamma = 2.1$.
The total mass inside the last measured point of the rotation curve is
$1.1 \times 10^{11} h_0^{-1} M_{\odot}$, which yields a total
mass-to-light-ratio $\Upsilon = 28 h_0 \Upsilon_{\odot}$. 
This has to be considered as a lower limit, since there is certainly
still a lot of dark matter beyond the last measured point on the
rotation curve.
The dark halo is at least four times as massive as the disk.
Such a value for the mass-to-light-ratio is typical for
spiral galaxies. Similar conclusions hold also for elliptical galaxies,
although one cannot measure rotation velocities.

\subsection{Mass of the Milky Way}

There are also measurements of the rotation velocity for our Galaxy.
However, these observations turn out to be rather difficult, and
the rotation curve has been measured only up to a distance of about
20 kpc. Without any doubt our own galaxy has a typical flat 
rotation curve.
A fact this which imply that it is possible to search directly for dark matter
characteristic of spiral galaxies in our own Milky Way.

In oder to infer the total mass one can also study the proper
motion of the Magellanic Clouds and of other satellites of our
Galaxy.
Recent studies \cite{kn:Zaritsky,kn:Lin,kn:Kochanek}
do not yet allow an accurate 
determination of $v_{rot}(LMC)/v_0$ 
($v_0 = 210 \pm 10$ km/s  being the local rotational velocity).
Lin et al. \cite{kn:Lin}  
analyzed the proper motion observations and concluded that 
within 100 kpc the Galactic halo has a mass 
$\sim 5.5 \pm 1 \times 10^{11} M_{\odot}$ and a substantial fraction 
$\sim 50\%$ of this mass is distributed beyond the present distance
of the Magellanic Clouds of about 50 kpc. Beyond 100 kpc the mass may 
continue to increase to $\sim 10^{12} M_{\odot}$ within its tidal radius
of about 300 kpc. This value for the total mass of the Galaxy is in
agreement with the results of Zaritsky et al. \cite{kn:Zaritsky}, who found
a total mass in the range 9.3 to 12.5 $\times 10^{11} M_{\odot}$, the former
value by assuming radial satellite orbits whereas the latter by assuming
isotropic satellite orbits.
 
The results of Lin et al. \cite{kn:Lin} suggest that
the mass of the halo dark matter up to the Large Magellanic Cloud
(LMC) is roughly half of the value
one gets for the standard halo model (with flat rotation
curve up to the LMC and spherical shape), implying thus the same reduction
for the number of expected microlensing events.
Kochanek \cite{kn:Kochanek} analysed the global mass distribution of the
Galaxy adopting a Jaffe model, whose parameters are determined
using the observations on the
proper motion of the satellites of the Galaxy, the Local
Group timing constraint and the ellipticity of the M31 orbit. 
With these observations Kochanek \cite{kn:Kochanek}
concludes that the mass inside 50 kpc is $5.4 \pm 1.3 \times 
10^{11} M_{\odot}$.
This value becomes, however, slightly smaller when using only the satellite 
observations and the disk rotation constraint, in this case
the median mass interior to 50 kpc is in the interval 3.3 to 6.1
(4.2 to 6.8) without (with) Leo I satellite in units of $10^{11} M_{\odot}$.
The lower bound without Leo I is 65\% of the mass expected assuming
a flat rotation curve up to the LMC.

\section{Neutrinos as halo dark matter}

For stable neutrinos (with mass $<$ 1 MeV) one gets the following
cosmological upper bound on the sum of their masses
\cite{kn:Gerstein,kn:Cowsik}
\begin{equation}
\sum_{\nu} m_{\nu} < 200h_0^2~  eV ~.
\end{equation}
If neutrinos make up the dark matter in the galactic halos, we may decribe
them as forming a bound system which resembles in the central regions
to an isothermal gas sphere.
The core radius of such an isothermal sphere is
\begin{equation}
r_c = \left ( \frac{9 \sigma^2}{4 \pi G \rho_0} \right)^{1/2}~,
\end{equation}
where $\sigma$ is the one-dimensional velocity dispersion and $\rho_0$
is the central density.
The velocity distribution of the neutrinos is Maxwellian
and the maximum phase-space density is
\begin{equation}
n_c =  \frac{4.5}{(2\pi)^5 G~ r_c^2~ \sigma~ m_{\nu}^4}~.
\end{equation}
The requirement that the maximum phase-space density does not violate
the Pauli exclusion principle ($n_c < g_{\nu}/h^3$, where
$g_{\nu}$ is the number of helicity states and $h$ Planck's constant)
leads then to the following
lower limit for the neutrino mass \cite{kn:Tremaine}
\footnote{
One gets a slightly higher bound, by a factor $2^{1/4}$,
using the fact that neutrinos behaves practically as collisionless
particles and thus by applying Liouville's theorem \cite{kn:Tremaine}.}
\begin{equation}
m_{\nu} > 120~ eV \left(\frac{100~ km ~s^{-1}}{\sigma} \right)^{1/4}
\left(\frac{1~ kpc}{r_c} \right)^{1/2} g_{\nu}^{-1/4}~.
\end{equation} 
Typical values for spiral galaxies are $\sigma \simeq 150$ km s$^{-1}$
and $r_c \simeq 20$ kpc. This way we get a lower bound
$m_{\nu} > 25 - 30$ eV \cite{kn:Tremaine,kn:Paganini},
which is still consistent
with the cosmological bound eq.(4).
See also refs. \cite{kn:Caz1,kn:Caz2} for a discussion of more
precise bounds for spirals and ellipticals by considering different
visible and dark matter distributions as well as the case where the
neutrinos are not fully degenerate.
However, when considering dwarf galaxies for which $r_c < 2$ kpc
and $\sigma \sim 10 - 30$ km s$^{-1}$ one gets
$m_{\nu} > 100 - 500$ eV \cite{kn:Linfaber}, which is clearly
in contraddiction with the cosmological bound, excluding thus
neutrinos as dark matter candidate for the halo of dwarf galaxies
and in turn of spiral galaxies.

This latter point follows also from considering the dwarf galaxies
Draco and Ursa Minor, which are both satellites of our Galaxy
\cite{kn:Gerhard} and, therefore, are located in its halo.
In fact, if their dark matter halo consist of neutrinos with
mass $\sim 30$ eV, then $r_c \sim 10$ kpc and the total mass
would be $\sim 4 \times 10^{11}~ M_{\odot}$.
However, such a high value for the total mass can be excluded
by the requirement that the dynamical friction time for such a satellite
galaxy moving in the halo of our Galaxy has to be longer
than the age of Galaxy $\sim 10^{10}$ yr. 
The upper value for the total mass one infers this way
is of the order of $10^{10}~M_{\odot}$.
Therefore, one gets
a lower limit of $\sim 80$ eV for the neutrino mass.

\section{Baryonic dark matter}

Before discussing the baryonic dark matter
we would like to mention that another
class of candidates which is seriously taken into consideration
is the so-called cold dark matter, which
consists for instance of axions
or supersymmetric particles like neutralinos \cite{kn:jungman}.
Here, we will not discuss cold dark matter
in detail. However, recent studies
seem to point out that there is a discrepancy between the calculated (through
N-body simulations)
rotation curve for dwarf galaxies assuming an halo of cold dark matter
and the measured curves \cite{kn:moore,kn:navarro}. 
If this fact is confirmed, this
would exclude cold dark matter as a major constituent of the 
halo of dwarf galaxies and possibly also of spiral
galaxies.

From the Big Bang nucleosynthesis model \cite{kn:copi,kn:PDG} 
and from the observed 
abundances of primordial elements one infers:
$0.010 \leq h^2_0 \Omega_B \leq 0.016$ or
with $h_0 \simeq 0.4 - 1$ one gets $0.01 \leq \Omega_B \leq 0.10$
(where $\Omega_B = \rho_B /\rho_{crit}$, and $\rho_{crit}=3H_0^2/8\pi G$).
Since for the amount of luminous baryons one finds
$\Omega_{lum} \ll \Omega_B$, it follows that an
important fraction 
of the baryons are dark.
In fact the dark baryons may well make up the entire dark halo matter.

The halo dark matter cannot be in the form of hot 
ionized hydrogen gas otherwise there would be a large
X-ray flux, for which there are stringent upper limits.
The abundance of neutral hydrogen gas
is inferred from the 21-cm measurements, which show that its contribution is 
small. Another possibility is that the hydrogen gas is in molecular form
clumped into cold clouds, as  
we will discuss in some detail in Section 6.
Baryons could otherwise have been processed in stellar remnants
(for a detailed discussion see \cite{kn:Carr}).
If their mass is below $\sim0.08~M_{\odot}$ they are too light to ignite
hydrogen burning reactions. 
The possible origin of such brown dwarfs or Jupiter like bodies
(called also MACHOs),
by fragmentation or by some other mechanism, is at present
not understood.  It has also been pointed out that the mass distribution
of the MACHOs, normalized to the dark halo mass density, could be 
a smooth continuation of the known initial mass function 
of ordinary stars
\cite{kn:Derujula1}. 
The ambient radiation, or their own body heat, would make
sufficiently small objects of H and He evaporate rapidly.
The condition that the rate of evaporation of such a hydrogenoid sphere be
insufficient to halve its mass in a billion years leads to the 
following lower limit on their mass \cite{kn:Derujula1}: 
$M > 10^{-7} M_{\odot}(T_S /30~ K)^{3/2} (1~ g~cm^{-3}/ \rho)^{1/2}$
($T_S$ being their surface
temperature and $\rho$ their average density, which we expect
to be of the order $\sim 1~ g~ cm^{-3}$).

Otherwise, 
MACHOs might be either M-dwarfs or else white dwarfs.
As a matter of fact, a deeper analysis shows that the M-dwarf option
look problematic. The null result of several searches for low-mass stars
both in the disk and in the halo of our
galaxy suggest that the halo cannot be mostly in the form of hydrogen
burning main sequence M-dwarfs. Optical imaging of high-latitude
fields taken with the Wide Field Camera of the Hubble Space Telescope
indicates that less than $\sim 6\%$ of the halo can be in this
form \cite{kn:Bahcall}. Also a substantial component of neutron
stars and black holes with mass higher than $\sim 1~M_{\odot}$ 
is excluded, for otherwise they would  lead to an overproduction of heavy 
elements relative to the observed abundances. A scenario
with white dwarfs as a major constituent of the galactic halo
dark matter has been explored \cite{kn:Tamanaha}.
However, it requires a rather ad hoc initial mass function sharply 
peaked around 2 - 6 $M_{\odot}$. Future
Hubble deep field exposures could either find the white dwarfs 
or put constraints on their fraction in the halo \cite{kn:Kawaler}.

\section{Detection of MACHOs through microlensing}

It has been pointed out by Paczy\'nski \cite{kn:Paczynski} that microlensing 
allows the detection of MACHOs located in the galactic halo in the mass
range \cite{kn:Derujula1}  
$10^{-7} < M/M_{\odot} <  1$.
In September 1993 the French collaboration EROS \cite{kn:Aubourg}
announced the discovery of 2 microlensing candidates
and the American--Australian
collaboration MACHO of one candidate \cite{kn:Alcock}.
In the meantime the MACHO team reported the observation of
altogether 8 events
(one of which is a binary lensing event) analyzing
2 years of their data
by monitoring about 8.5 million 
of stars in the LMC \cite{kn:Pratt}. 
Their analysis leads to an optical depth of $\tau=2.9^{+1.4}_{-0.9}
\times 10^{-7}$ and correspondingly to a halo MACHO fraction of the
order of 45-50\% and an average mass $0.5^{+0.3}_{-0.2} M_{\odot}$, 
under the assumption of a standard spherical halo model.
It may well be that there is also a contribution of events due
to MACHOs located in the LMC itself or in a thich disk of our galaxy,
the corresponding optical depth is estimated to be
$\tau=5.4 \times 10^{-8}$
\cite{kn:Pratt}.
EROS has also searched for very-low mass MACHOs by looking for
microlensing events with time scales ranging from 30 minutes to 
7 days \cite{kn:EROS}. The lack of candidates in this range 
places significant constraints on any model for the halo that relies
on objects in the range $5 \times 10^{-8} < M/M_{\odot} < 5 \times 10^{-4}$.
Similar conclusions have been reached also by the MACHO team
\cite{kn:Pratt}. 
Moreover, the
Polish-American team OGLE \cite{kn:Udalski}, the MACHO  
\cite{kn:MACHO} and the French DUO \cite{kn:Alard} collaborations 
found altogether more than $\sim$ 100  
microlensing events by monitoring stars located in the galactic bulge.
The inferred optical depth for the bulge turns out to be
higher than previously thought. These results are very important
in order to study the structure of our Galaxy.

In the following we present the main features of microlensing,
in particular its probability and the rate of events \cite{kn:Roulet}.
An important issue is 
the determination from the observations of the mass of the MACHOs that
acted as gra\-vi\-tational lenses as well as the fraction of halo dark
matter they make up.
The most appropriate way to compute the average mass and other
important information is to use
the method of mass moments developed by De R\'ujula et al. \cite{kn:Derujula},
which will be briefly discussed in Section 5.4.

\subsection{Microlensing probability}

When a
MACHO of mass $M$ is sufficiently close to the line of sight
between us and a more distant
star, the light from the source suffers a gravitational
deflection (see Fig. 2). 
The deflection angle is usually so small that we do not see
two images but rather a magnification  of the original star brightness.
This magnification, at its maximum, is given by
\begin{equation}
A_{max}=\frac{u^2+2}{u(u^2+4)^{1/2}}~ . \label{eq:bb}
\end{equation}
Here $u=d/R_E$ ($d$ is the distance of the MACHO from the line of sight)
and the Einstein radius $R_E$ is defined as
\begin{equation}
R_E^2=\frac{4GMD}{c^2}x(1-x) \label{eq:cc}
\end{equation}
with $x=s/D$, and
where $D$ and $s$ are the distance between the source, respectively 
the MACHO and the observer. 

An important quantity is the optical depth $\tau_{opt}$ 
to gravitational microlensing defined as
\begin{equation}
\tau_{opt}=\int_0^1 dx \frac{4\pi G}{c^2}\rho(x) D^2 x(1-x)
\label{eq:za}
\end{equation}
with $\rho(x)$ the mass density of microlensing matter at distance
$s=xD$ from us along the line of sight. 
The quantity $\tau_{opt}$ is the probability
that a source is found within a radius $R_E$ of some MACHO and thus has a
magnification that is larger
than $A= 1.34$ ($d \leq R_E$).

We calculate $\tau_{opt}$ for a galactic mass
distribution of the form
\begin{equation}
\rho(\vec r)=\frac{\rho_0(a^2+R^2_{GC})}
{a^2+\vec r^2}~, \label{eq:zb}
\end{equation}
$\mid \vec r \mid$ being the distance from the Earth.
Here, $a$ is the core radius,
$\rho_0$ the local dark mass
density in the solar system and $R_{GC}$ the distance
between the observer and the Galactic centre.
Standard values for the
parameters are
$\rho_0=0.3~GeV/cm^3=7.9~10^{-3} M_\odot/pc^3$,
$a=5.6~kpc$ and $R_{GC}=8.5~kpc$.
With these values we get, for a spherical halo, $\tau_{opt}=0.7 \times 10^{-6}$
for the LMC and $\tau_{opt}=10^{-6}$ for the SMC \cite{kn:locarno}.

The magnification of the brightness of a star by a MACHO is a time-dependent
effect.
For a source that can be considered as
pointlike (this is the case if the projected star radius at the MACHO
distance is much less than $R_E$) 
the light curve as a function of time is obtained by inserting
\begin{equation}
u(t)=\frac{(d^2+v^2_Tt^2)^{1/2}}{R_E} \label{eq:zd}
\end{equation}
into eq.(\ref{eq:bb}), 
where $v_T$ is the transverse velocity of the MACHO, which can be inferred
from the measured rotation curve ($v_T \approx 200~ km/s$). The
achromaticity, symmetry and uniqueness of the signal are distinctive
features that allow to discriminate a microlensing event from
background events such as variable stars.

The behaviour of the magnification with time, $A(t)$, determines two
observables namely, the magnification at the peak $A(0)$ - denoted
by $A_{max}$ -
and the width of the signal $T$ (defined as 
being $T = R_E/v_T$).

\subsection{Microlensing rates}

The microlensing rate depends on the mass and velocity distribution of
MACHOs. 
The mass density at a distance $s=xD$ from the observer is given by
eq.(\ref{eq:zb}).
The isothermal
spherical halo model does not determine the MACHO number density as a
function of mass. A
simplifying  assumption is to let the mass distribution be independent
of the position in the galactic halo, i.e., we assume the following
factorized form for the number density per unit mass $dn/dM$,
\begin{equation}
\frac{dn}{dM}dM=\frac{dn_0}{d\mu}
\frac{a^2+R_{GC}^2}{a^2+R_{GC}^2+D^2x^2-2DR_{GC}x cos\alpha}~d\mu=
\frac{dn_0}{d\mu} H(x) d\mu~,
\label{eq:zj}
\end{equation}
with $\mu=M/M_{\odot}$ ($\alpha$ is the angle of the 
line of sight with 
the direction of the galactic centre), $n_0$ not depending on $x$ 
and is subject to the normalization
$\int d\mu \frac{dn_0}{d\mu}M=\rho_0$.
Nothing a priori is known on the distribution $d n_0/dM$.

A different situation arises for the velocity
distribution in the isothermal
spherical halo model, its
projection in the plane perpendicular to the line of sight
leads to the following
distribution in the transverse velocity $v_T$
\begin{equation}
f(v_T)=\frac{2}{v_H^2}v_T e^{-v^2_T/v_H^2}.\label{eq:zr}
\end{equation}
($v_H \approx 210~km/s$ is the observed velocity dispersion in the halo).

In order to find the rate at which a single star
is microlensed with magnification
$A \geq A_{min}$, we consider MACHOs
with masses between $M$ and $M+\delta M$, located at a distance from
the observer between $s$ and $s+\delta s$ and with transverse velocity
between $v_T$ and $v_T+\delta v_T$. The collision time can be
calculated using the well-known fact that the inverse of the collision
time is the product of the MACHO number density, the microlensing
cross-section and the velocity. 
The rate $d\Gamma$, taken also as a differential with respect 
to the variable $u$, at which a single star is microlensed
in the interval $d\mu du dv_T dx$ is given by
\cite{kn:Derujula,kn:Griest1}
\begin{equation}
d\Gamma=2v_T f(v_T)D r_E [\mu x(1-x)]^{1/2} H(x)
\frac{d n_0}{d\mu}d\mu du dv_T dx,\label{eq:zt}
\end{equation}
with
\begin{equation}
r_E^2=\frac{4GM_{\odot}D}{c^2} \sim
(3.2\times 10^9 km)^2 .\label{eq:zs}
\end{equation}

One has to integrate
the differential number of microlensing events, 
$dN_{ev}=N_{\star} t_{obs} d\Gamma$,
over an appropriate range for $\mu$, $x$,
$u$ and $v_T$, 
in order to obtain the total number of microlensing events which can
be compared with an experiment
monitoring $N_{\star}$ stars during an
observation time $t_{obs}$ and which is able to detect
a magnification such that $A_{max} \geq A_{TH}$.
The limits of the $u$ integration are determined by
the experimental threshold in magnitude shift, $\Delta m_{TH}$:
we have $0 \leq u \leq u_{TH}$.

The range of integration for $\mu$ is where the mass
distribution $dn_0/d\mu$ is not vanishing
and that for $x$ is
$0\leq x \leq D_h/D$ where $D_h$ is the extent of the galactic halo along
the line of sight (in the case of the LMC,
the star is inside the galactic halo and thus $D_h/D=1$.)
The galactic velocity distribution is cut at the escape velocity
$v_e \approx 640~km/s$ and therefore
$v_T$ ranges over $0\leq v_T \leq v_e$.
In order to simplify the integration we integrate $v_T$
over all the positive axis, due to the exponential factor in $f(v_T)$ the
so committed error is negligible.

However, the integration range of $d\mu du dv_T dx$
does not span all the interval we have just described.
Indeed, each experiment has time
thresholds $T_{min}$ and $T_{max}$ and only detects events with:
$T_{min}\leq T \leq T_{max}$,
and thus the integration range has to be such that $T$ lies in this
interval.
The total number of micro-lensing events is then given by
\begin{equation}
N_{ev}=\int dN_{ev}~\Theta (T-T_{min})\Theta (T_{max}-T),\label{eq:th}
\end{equation}
where the integration is over the full range of
$d\mu du dv_T dx$. $T$ is related in a complicated way
to the integration variables,
because of this, no direct
analytical integration in eq.(\ref{eq:th}) can be performed.

To evaluate eq.(\ref{eq:th}) we define
an efficiency function $\epsilon_0(\mu)$
which measures the fraction of the total number of microlensing events
that meet the condition on $T$ at a
fixed MACHO mass $M=\bar\mu M_{\odot}$.
A more detailed analysis \cite{kn:Derujula} 
shows that
$\epsilon_0(\mu)$
is in very good approximation equal to unity for possible MACHO objects
in the mass
range of interest here.
We now can write the total number of events in
eq.(\ref{eq:th}) as
\begin{equation}
N_{ev}=\int dN_{ev}~\epsilon_0(\mu).\label{eq:tl}
\end{equation}
Due to the fact that
$\epsilon_0$ is a function of $\mu$ alone, the integration in
$d\mu du dv_T dx$ factorizes into four integrals with independent
integration limits. 

In order to quantify the expected number of events it is convenient
to take as an example a delta function distribution for the mass.
The rate of microlensing
events with
$A \geq A_{min}$ (or $u \leq u_{max}$), is then
\begin{equation}
\Gamma(A_{min})=\tilde\Gamma u_{max}= 
D r_E u_{max} \sqrt{\pi}~v_H \frac{\rho_0}{M_{\odot}}\frac{1}{\sqrt{\bar \mu}}
\int^1_0 dx[x(1-x)]^{1/2} H(x)~.\label{eq:ta}
\end{equation}

Inserting the numerical values for the LMC
(D=50~kpc and $\alpha=82^0$) we get
\begin{equation}
\tilde\Gamma=4
\times 10^{-13}~\frac{1}{s}~\left( \frac{v_H}{210~km/s}\right)
 \left(\frac{1}{\sqrt{D/kpc}}\right)
 \left( \frac{\rho_0}{0.3~GeV/cm^3}\right)
\frac{1}{\sqrt{M/M_{\odot}}}\ .
\label{eq:tb}
\end{equation}
For an experiment monitoring $N_{\star}$ stars during an
observation time $t_{obs}$ the total number of events with a
magnification $A \geq A_{min}$ is:
$N_{ev}(A_{min})=N_{\star} t_{obs} \Gamma(A_{min})$.
In the following Table 1 we show some values of $N_{ev}$ for the LMC,
taking
$t_{obs}=10^7$ s ($\sim$ 4 Months), $N_{\star}=10^6$ stars and 
$A_{min} = 1.34$ (or $\Delta m_{min} = 0.32$).
\vskip 0.3 cm 
Table 1

\begin{center}
\begin{tabular}{|c|c|c|c|}\hline
MACHO mass in units 
of $M_{\odot}$ & Mean $R_E$ in km & Mean microlensing time &
$N_{ev}$ \\
\hline
$10^{-1}$ & $0.3\times 10^9$ & 1 month & 1.5  \\
$10^{-2}$ & $10^8$ & 9 days & 5 \\
$10^{-4}$ & $10^7$ & 1 day & 55 \\
$10^{-6}$ & $10^6$ & 2 hours & 554 \\
\hline
\end{tabular}
\end{center}

Gravitational microlensing could also be useful for detecting MACHOs in
the halo of nearby galaxies \cite{kn:Crotts,kn:Baillon} 
such as M31 or M33, for which experiments are under way.
In fact, it turns out
that the massive dark halo of M31 has an optical depth to microlensing
which is of about the same order of magnitude as that of our own
galaxy $\sim 10^{-6}$ \cite{kn:Crotts,kn:Jetzer}. 
Moreover, an experiment monitoring stars in
M31 would be sensitive to both MACHOs in our halo and in the one of M31.
One can also compute the microlensing rate \cite{kn:Jetzer}
for MACHOs in the halo of M31, for which we get 
\begin{equation}
\tilde\Gamma=1.8 \times 10^{-12} \frac{1}{s} \left(\frac{v_H}{210~km/s} \right)
\left(\frac{1}{\sqrt{D/kpc}}\right)\left(\frac{\rho(0)}{1~Gev/cm^3} \right)
\frac{1}{\sqrt{M/M_{\odot}}}~. \label{eq:tc}
\end{equation}
($\rho(0)$ is the central density of dark matter.)
In the following Table 2 we show some values of $N^a_{ev}$ due to MACHOs in the
halo of M31 with $t_{obs}= 10^7$ s and $N_{\star}=10^6$ stars. In the 
last column we give the corresponding number of events, $N_{ev}$,
due to MACHOs in our    
own halo. The mean microlensing time is about the same for both types of 
events.
\vskip 0.3cm 
Table 2

\begin{center}
\begin{tabular}{|c|c|c|c|c|}\hline
MHO mass in units of $M_{\odot}$ & Mean $R_E$ in km & Mean microlensing time &
$N^a_{ev}$ & $N_{ev}$\\
\hline
$10^{-1}$ & $7\times 10^8$ & 38 days & 2 & 1 \\
$10^{-2}$ & $2\times 10^8$ & 12 days & 7 & 4 \\
$10^{-4}$ & $2\times 10^7$ & 30 hours & 70 & 43 \\
$10^{-6}$ & $2\times 10^6$ & 3 hours & 700 & 430 \\      
\hline
\end{tabular}
\end{center}

Of course these
numbers should be taken as an estimate, since they depend on the details 
of the model one adopts for the distribution of the dark matter in the halo.\\

\subsection{Most probable mass for a single event}

The probability $P$ that a microlensing
event of duration T
and maximum amplification $A_{max}$ be produced by a MACHO
of mass $\mu$ (in units of $M_{\odot}$) is given by
\cite{kn:Jetzer2} 
\begin{equation}
P(\mu,T) \propto \frac{\mu^2}{ T^4} \int_0^1 dx (x(1-x))^2 H(x)
exp\left( -\frac{r_E^2 \mu x(1-x)}{v^2_H  T^2} \right) ~, \label{eqno:8}
\end{equation}
which does not dependent on $A_{max}$ 
and $P(\mu, T)=P(\mu/ T^2)$. The measured values for
$ T$ are listed in Table 3, where 
$\mu_{MP}$ is
the most probable value.
We find that the maximum corresponds to 
$\mu r_E^2/v^2_H T^2=13.0$ \cite{kn:Jetzer2,kn:Jetzer1}. 
The 50\% confidence interval
embraces for the mass $\mu$ approximately
the range $1/3\mu_{MP}$ up to $3 \mu_{MP}$.
Similarly one can compute $P(\mu, T)$ also for the bulge events
(see \cite{kn:Jetzer1}).

\vskip 0.5cm 
Table 3: Values of $\mu_{MP}$ (in $M_{\odot}$)
for eight microlensing events detected in the LMC ($A_{i}$
= American-Australian
collaboration events ($i$ = 1,..,6);
$F_1$ and $F_2$
French collaboration events).
For the LMC: $v_H = 210~{\rm km}~{\rm s}^{-1}$ and
$r_E = 3.17 \times 10^9~{\rm km}$.

\begin{center}
\begin{tabular}{|c|c|c|c|c|c|c|c|c|}\hline
  & $A_{1}$ & $A_{2}$ & $A_3$ & $A_4$ & $A_5$ & $A_6$ & $F_1$ & $F_2$  \\
\hline
$ T$ (days) & 17.3 & 23 & 31 & 41 & 43.5 & 57.5 & 27 & 30 \\
\hline
$\tau (\equiv \frac{v_H}{r_E} T)$ & 0.099 &
0.132 & 0.177 & 0.235 & 0.249 & 0.329 & 0.155 & 0.172 \\
\hline
$\mu_{MP}$ & 0.13 & 0.23 & 0.41 & 0.72 & 0.81 & 1.41 & 0.31 & 0.38 \\
\hline
\end{tabular}
\end{center} 

\subsection{Mass moment method}

A more systematic way to extract information on the masses is to use the
method of mass moments \cite{kn:Derujula}. 
The mass moments $<\mu^m>$ are defined as
\begin{equation}
<\mu^m>=\int d\mu~ \epsilon_n(\mu)~ 
\frac{dn_0}{d\mu}\mu^m~. \label{eqno:10}
\end{equation}
$<\mu^m>$ is related to $<\tau^n>=\sum_{events} \tau^n$,
with $\tau \equiv (v_H/r_E) T$, as constructed
from the observations and which can also be computed as follows
\begin{equation}
<\tau^n>=\int dN_{ev}~ \epsilon_n(\mu)~
\tau^n=V u_{TH} \Gamma(2-m) \widehat H(m) <\mu^m>~,
\label{eqno:11}
\end{equation}
with $m \equiv (n+1)/2$ and
\begin{equation}
V \equiv 2 N_{\star} t_{obs}~ D~ r_E~ v_H=2.4 \times 10^3~ pc^3~ 
\frac{N_{\star} ~t_{obs}}{10^6~ {\rm star-years} }~, \label{eqno:12}
\end{equation}
\begin{equation}
\Gamma(2-m) \equiv \int_0^{\infty} \left(\frac{v_T}{v_H}\right)^{1-n}
f(v_T) dv_T~,
\label{eqno:13}
\end{equation}
\begin{equation}
\widehat H(m) \equiv \int_0^1 (x(1-x))^m H(x) dx~.  \label{eqno:14}
\end{equation}
The efficiency $\epsilon_n(\mu)$ is determined as follows 
(see \cite{kn:Derujula})
\begin{equation}
\epsilon_n(\mu) \equiv \frac{\int d N^{\star}_{ev}(\bar\mu)~ 
\epsilon(T)~ \tau^n}
{\int d N^{\star}_{ev}(\bar\mu)~ \tau^n}~, \label{eqno:15}
\end{equation}
where $d N^{\star}_{ev}(\bar\mu)$ is defined as $d N_{ev}$ 
in eq.(\ref{eq:th}) with
the MACHO mass distribution concentrated at a fixed mass
$\bar\mu$: $dn_0/d\mu=n_0~ \delta(\mu-\bar\mu)/\mu$. 
$\epsilon(T)$ is the experimental detection efficiency.
For a more detailed discussion on the efficiency see ref. \cite{kn:Masso}.

A mass moment $< \mu^m >$ is thus related to 
$< \tau^n >$ as given from the measured values 
of $T$ in a microlensing experiment by
\begin{equation}
< \mu^m > = \frac{< \tau^n >}{V u_{TH} \Gamma(2-m) \hat H(m)}~.
\label{eqno:16}
\end{equation}

The mean local density of MACHOs (number per cubic parsec)
is $<\mu^0>$. The average local mass density in MACHOs is
$<\mu^1>$ solar masses per cubic parsec. 
In the following we consider only 6 (see Table 3)
out of the 8 events observed by the MACHO group,
in fact the two events we neglect are 
a binary lensing event and an event which is rated as marginal.
The  mean mass, which we get from
the six events detected by the MACHO team, is 
\begin{equation}
\frac{<\mu^1>}{<\mu^0>}=0.27~M_{\odot}~.
\label{eqno:aa}
\end{equation}
(To obtain this result we used the values of $\tau$
as reported in Table 3, whereas $\Gamma(1)\widehat H(1)=0.0362$ and
$\Gamma(2)\widehat H(0)=0.280$ as
plotted in Fig. 6 of ref. \cite{kn:Derujula}).
If we include also the two EROS events we get a value
of 0.26 $M_{\odot}$ for the mean mass.
The resulting mass depends on the parameters
used to describe the standard halo model. In order to check this
dependence we varied the parameters within
their allowed range and found
that the average mass changes at most by $\pm$ 30\%, which shows
that the result is rather robust. 
Although the value for the average mass we find with the mass moment
method is marginally consistent with the result of the MACHO team,
it definitely favours a lower average MACHO mass.

One can
also consider other models with more general
luminous and dark matter distributions, e.g. ones with a flattened halo
or with anisotropy in velocity space \cite{kn:Ingrosso},
in which case the resulting
value for the average mass would decrease significantly.
If the above value will be confirmed, then MACHOs cannot be brown dwarfs
nor ordinary hydrogen burning stars, since for the latter
there are observational
limits from counts of faint red stars. Then white dwarfs
are the most likely explanation. As mentioned 
in Section 4 such a scenario 
has been explored recently \cite{kn:Tamanaha}.
However, it has some problems, since it requires that
the initial mass function
must be sharply peaked around $2 - 6~ M_{\odot}$.
Given these facts, we feel that the brown dwarf option can still provide a 
sensible explanation of the observed microlensing events \cite{kn:DGJR}.

Another important quantity to be determined is the fraction $f$ of the local
dark mass density (the latter one given by $\rho_0$) detected
in the form of MACHOs, which is given by
$f \equiv {M_{\odot}}/{\rho_0} \sim 126~{\rm pc}^3$ $<\mu^1>$.
Using the values given by the MACHO collaboration
for their two years data \cite{kn:Pratt} (in particular
$u_{TH}=0.661$ corresponding to $A > 1.75$ and
an effective exposure $N_{\star} t_{obs}$
of $\sim 5 \times 10^6$ star-years for 
the observed range of the event duration $T$ between $\sim$ 20 - 50 days)
we find $f \sim 0.54$, which compares quite well
with the corresponding value ($f \sim 0.45$ based on the six events
we consider) calculated
by the MACHO group in a different way. The value for $f$ is obtained 
again by assuming a standard spherical halo model. 

\newpage
Table 4: Values of
$\mu_{MP}$ (in $M_{\odot}$) as
obtained by the corresponding $P(\mu, T)$ for 
eleven microlensing events detected by OGLE
in the galactic bulge \cite{kn:Masso}. 
($v_H = 30~{\rm km}~{\rm s}^{-1}$ and
$r_E = 1.25 \times 10^9~{\rm km}$.) ($T$ is in days as above.)

\begin{center}
\begin{tabular}{|c|c|c|c|c|c|c|c|c|c|c|c|}\hline
  & 1 & 2 & 3 & 4 & 5 & 6 & 7 & 8 & 9 & 10 & 11\\
\hline
$ T$ & 25.9 & 45 & 10.7 & 14 &12.4& 8.4& 49.5&18.7&61.6&12&20.9 \\
\hline
$\tau$ & 0.054 &
0.093 & 0.022 & 0.029 & 0.026&0.017& 0.103& 0.039& 0.128& 0.025& 0.043\\
\hline
$\mu_{MP}$ & 0.61 & 1.85 & 0.105 & 0.18 &0.14& 0.065& 2.24& 0.32&
3.48 & 0.13 & 0.40 \\
\hline
\end{tabular}
\end{center}
\vskip 0.2cm

Similarly, one can also get information from the events
detected so far towards the galactic bulge.
The mean MACHO mass, which one gets when considering
the first eleven events detected by OGLE in the galactic bulge (see Table 4),
is $\sim 0.29 M_{\odot}$ \cite{kn:Jetzer1}.
From the 40 events discovered 
during the first year of operation
by the MACHO team \cite{kn:MACHO} (we considered
only the events used by the MACHO team to infer the optical
depth without the double lens event)
we get an average value
of 0.16$M_{\odot}$.
The lower value inferred from the MACHO data is due to the fact
that the efficiency for the short duration events ($\sim$ some days)
is substantially higher for the MACHO experiment than for the
OGLE one. 
These values for the average mass
suggest that the lens are faint disk stars. 

Once several moments $< \mu^m >$ are known one can
get information on the mass distribution $dn_0/d\mu$. 
Since at present only few events towards the LMC are at disposal the 
different moments (especially the higher ones) can 
be determined only approximately.
Nevertheless, the results obtained so far
are already of interest and it is clear that in a few years,
due also to the new experiments under way (such as EROS II and OGLE II),
it will be possible to draw more firm conclusions.

\section{Dark clusters of MACHOs and cold molecular clouds}
 
A major problem which arises is to explain the formation of MACHOs, as well
as the nature of the remaining amount of dark matter in the galactic halo.
We feel it hard to conceive a formation mechanism which transforms with 100\%
efficiency hydrogen and helium gas into MACHOs. Therefore, we expect
that also cold clouds (mainly of $H_2$) should be present in the 
galactic halo. Recently, we have proposed
a scenario \cite{kn:Roc,kn:Roncadelli}
in which dark clusters of MACHOs and cold molecular 
coulds naturally form in the halo at galactocentric distances
larger than 10-20 kpc, where the relative abundance depends on the
distance (similar ideas have also been developed in refs.
\cite{kn:Pfenniger,kn:Silk}). Our scenario  
can be summarized as follows.

After its initial collapse, the proto galaxy (PG) is expected to be shock
heated to its virial temperature $\sim 10^6$ K. Since overdense regions cool
more rapidly than average (by hydrogen recombination), proto globular cluster
(PGC) clouds form in
pressure equilibrium with diffuse gas. At this stage, the PGC 
cloud temperature is $\sim 10^4$ K, its mass and size are
$\sim 10^6 (R/kpc)^{1/2} M_{\odot}$ and $\sim 10~(R/kpc)^{1/2}$ pc,
respectively.
The subsequent evolution of the PGC
clouds will be different in the inner and outer part of the galaxy, depending
on the decreasing collision rate and ultraviolet (UV) fluxes 
as the galactocentric distance increases.
Below $10^4$ K, the main coolants are $H_2$ molecules and any heavy element
produced in a first chaotic galactic phase.
In the central region of the galaxy
an Active Galactic Nucleus and/or a first population of  
massive stars are expected to exist,
which act as strong sources of UV radiation that dissociates
the $H_2$ molecules present in the inner part of the halo. As a consequence,
cooling is heavily suppressed and so inner PGC clouds remain for a long
time at temperature $\sim 10^4$ K, resulting in the imprinting
of a characteristic mass $\sim 10^6 M_{\odot}$.
Later on, the cloud temperature suddenly drops below
$10^4$ K and the subsequent evolution leads to the formation of stars 
and ultimately to stellar globular clusters.
In the outer regions of the halo 
the UV-flux is suppressed,
so that no substantial $H_2$ depletion actually happens. 
This fact has three distinct implications:
$(i)$ no imprinting of a characteristic PGC cloud mass shows up,
$(ii)$ the Jeans mass can now be lower than $10^{-1} M_{\odot}$,
$(iii)$  the cooling time is much shorter than the collision time.
PGC clouds subsequently fragment into 
smaller clouds that remain optically thin until the minimum value
of the Jeans mass is attained, thus 
leading to MACHO formation in dark clusters. 
Moreover, because the
conversion efficiency of the constituent gas in MACHOs 
could scarcely have been 100\%,
we expect the remaining fraction of the gas to form self-gravitating
molecular clouds, since, in the absence of 
strong stellar winds, the surviving gas
remains bound in the dark cluster, but not in diffuse form as in this case 
the gas would be observable in the radio band.
 
\subsection{Observational Tests} 
Let us now address the possible signatures of the above scenario, in addition 
to the single MACHO detection via microlensing.

We proceed to estimate the $\gamma$-ray flux produced in 
molecular clouds through the interaction
with high-energy cosmic-ray protons. 
Cosmic rays scatter on protons in the molecules producing $\pi^0$'s, 
which subsequently decay into $\gamma$'s. 
An essential ingredient is the knowledge of the cosmic ray
flux in the halo. Unfortunately, this quantity is experimentally unknown
and the only available information comes from theoretical estimates.
More precisely, from the mass-loss rate of a 
typical galaxy we infer a total cosmic ray flux in the halo
$F \simeq 1.1\times 10^{-4}$ erg cm$^{-2}$ s$^{-1}$.
We also need the energy distribution of the
cosmic rays, for which we assume the same energy dependence
as measured on the Earth. We then scale the overall
density in such a way that the integrated energy flux agrees with the above
value. Moreover, we assume that the cosmic ray density scales as
$R^{-2}$ for large galactocentric distance $R$. 
Accordingly, we obtain \cite{kn:Roc,kn:Roncadelli}
\begin{equation}
\Phi_{CR}(E, R) \simeq 1.9\times 10^{-3}~\Phi_{CR}^{\oplus}(E)~ 
\frac{a^2+R_{GC}^2}{a^2+R^2}~, \label{eqno:45}
\end{equation}
where $\Phi_{CR}^{\oplus}(E)$ is the measured primary cosmic ray flux on 
the Earth,
$a\sim 5$ kpc is the halo core radius and $R_{GC}\sim 8.5$ kpc is our
distance from the galactic center. 
The source function 
$q_{\gamma}(r)$, which gives the photon number density at distance
$r$ from the Earth, is 
\begin{equation}
q_{\gamma}(r)=\frac{4\pi}{m_p}\rho_{H_2}(r) \int dE_p~ \Phi_{CR}(E_p,R(r))~ 
\sigma_{in}(p_{lab}) <n_{\gamma}(E_p)>~. \label{eqno:49}
\end{equation}
Actually, the cosmic ray protons in the halo
which originate from the galactic disk are mainly directed outwards. This 
circumstance implies that the induced photons will predominantly 
leave the galaxy.
However, the presence of magnetic fields in the halo might give rise
to a temporary confinement of the cosmic ray protons similarly to what 
happens in the disk. 
In addition, there could also be sources of
cosmic ray protons located in the halo itself, as for instance
isolated or binary pulsars in globular clusters.
As we are unable to give a quantitative estimate of the 
above effects, we take them into account by introducing an
efficiency factor $\epsilon$, which could be rather small. In this way, 
the $\gamma$-ray photon flux reaching
the Earth is obtained by multiplying 
$q_{\gamma}(r)$ by $\epsilon/4\pi r^2$ and integrating
the resulting quantity over the cloud volume along the line of sight. 

The best chance to detect the $\gamma$-rays in question is provided
by observations at high galactic latitude. Therefore we 
find
\begin{equation}
\Phi_{\gamma}(90^0) \simeq \epsilon f~ 3.5 \times 10^{-6}~
{\rm \frac{photons}{cm^2~ s~sr}}~. \label{eqno:53}
\end{equation}

The inferred upper bound for $\gamma$-rays in the 0.8 - 6 GeV
range at high galactic latitude is $3 \times 10^{-7}$ photons cm$^{-2}$
s$^{-1}$ sr$^{-1}$ {\cite{kn:Bouquet}}.
Hence, we see from eq. (\ref{eqno:53}) that the presence of
halo molecular clouds does not lead nowadays to
any contradiction with such an upper limit,
provided $\epsilon f < 10^{-1}$.

Molecular clouds can be detected via the anisotropy they would introduce 
in the Cosmic Background Radiation (CBR), even if 
the ratio of the temperature excess 
of the clouds to the CBR temperature is less than $\sim 10^{-3}$. 
Consider molecular clouds in M31.
Because we expect they have
typical rotational speeds of 
$50~-~100$ km s$^{-1}$, the Doppler shift effect
will show up as an anisotropy in the CBR. The
corresponding anisotropy is then \cite{kn:dijqr}
\begin{equation}
\frac{\Delta T}{T_r}= \pm \frac{v}{c}~S~f~\tau_{\nu}~, \label{1}
\end{equation}
where $S$ is the spatial filling factor 
and $T_r$ is the CBR temperature.
If the clouds are optically thick only at some frequencies, one can use 
the average optical depth over the frequency range of the detector $\bar \tau$. 
We estimate the expected CBR anisotropy between two fields
of view (on opposite sides of M31) separated by $\sim 4\degr$
and with angular resolution of $\sim 1\degr$.
Supposing that the halo of M31 consists of
$\sim 10^6$ dark clusters and that all of them lie between 25 kpc and 35 kpc,
we would be able to detect $10^{3}-10^{4}$ dark clusters per degree
square. Scanning an  annulus of $1\degr$ width and internal
angular diameter $4\degr$, centered at M31, in 180 steps of $1\degr$, we
would find anisotropies of $\sim 2 \times 10^{-5}~f~\bar \tau$ in
$\Delta T/T_r$ (as now $S=1/25$). In conclusion, the theory does not
permit to establish whether the expected anisotropy lies above or below
current detectability ($\sim 10^{-6}$), and so only observations 
can resolve this issue.

An attractive strategy to discover the halo molecular clouds clumped
into dark clusters relies upon the absorption lines they would introduce in the
spectrum of a LMC star \cite{kn:PP}.
 
Let us now turn to the possibility of detecting MACHOs in M31 via their
infrared emission. For simplicity, we assume all MACHOs
have equal mass $\sim 0.08~M_{\odot}$ (which is the upper
mass limit for brown dwarfs) and make up the fraction $f$ of the
dark matter in M31. In addition, we suppose that all MACHOs have
the same age $t\sim 10^{10}$ yr {\cite{kn:ad}}. 
As a consequence, MACHOs emit most of their
radiation at the wavelength $\lambda_{max}\sim 2.6~{\rm \mu m}$.
The infrared surface brightness $I_{\nu}(b)$ of the M31 dark halo 
as a function of the projected separation $b$ (impact parameter) 
is given by
\begin{equation}
I_{\nu}(b) \sim  5\times 10^{5} \frac{x^3}{e^x-1}
\frac{a^2~f}{D\sqrt{a^2+b^2}} \arctan \sqrt { \frac{L^2-b^2} {a^2+b^2} }~~
{\rm Jy~sr^{-1}}, \label{6}
\end{equation}
where the M31 dark halo radius is taken to be $L \sim 50$ kpc.
Some numerical values of
$I_{\nu_{max}}(b)$ with $b=20$ and $40$ kpc are
$\sim 1.6 \times 10^{3}~f~{\rm Jy~sr^{-1}}$ and
$\sim 0.4 \times 10^{3}~f~{\rm Jy~sr^{-1}}$, respectively.
The planned SIRTF Satellite contains an
array camera with expected sensitivity of $\sim 1.7 \times 10^{3}~
{\rm Jy~sr^{-1}}$
per spatial resolution element in the wavelength range 2-6 $\mu$m.
Therefore, the MACHOs in the halo of M31 can, 
hopefully, be detected in the near future.

\section{Conclusions}

The mistery of the dark matter is still unsolved, however, thanks
to the ongoing microlensing experiments there is hope that
progress on its nature in the galactic halo
can be achieved within the next few years. 
It is well plausible that only a fraction of the halo
dark matter is in form of MACHOs, either brown dwarfs or white
dwarfs, in which case there is the problem of explaining the nature
of the remaining dark matter and the formation of the MACHOs.
Before invoking the need for new particles
as galactic dark matter candidates
for the remaining fraction, one should seriously consider the
possibility that it is in the form of cold molecular clouds.
A scenario this, for which several observational tests
have been proposed, thanks to which it should be feasible
in the near future 
to either detect or to put stringent limits on these clouds. 

I would like to thank F. De Paolis for carefully reading the manuscript.

\begin{figcap}
\item
Rotation curve for NGC 3198 according to van Albada et al. \cite{kn:vanAlbada}.
The dotted line with error bars refers to the optical and 21 cm  hydrogen data,
while the solid lines are theoretical fits.

\item Definition of various quantities describing a microlensing event.
The observer is O, the source is S and M is the MACHO.




\end{figcap}

\end{document}